\documentclass[11pt]{article}
\usepackage{amsmath,epsfig,graphicx,amsbsy,amssymb,citesort,url}
\usepackage{algorithm,algorithmic,fullpage}

\def\real    { \mathbb{R} }
\def \M {\mathcal{M}}

\def \R {\mathcal{R}}
\def \P {\mathcal{P}}
\def \C {\mathcal{C}}
\def \S {\mathcal{S}}
\def \A {\mathcal{A}}
\def \B {\mathcal{B}}
\def \trans {^T} 

\def \btheta {\mathbf{\theta}}
\def \bPhi {\mathbf{\Phi}}
\def \bPsi {\mathbf{\Lambda}}
\def \bpsi {\mathbf{\lambda}}

\def \x {{\bf x}}
\def \a {{\bf a}}
\def \b {{\mathbf b}}
\def \g {{\mathbf g}}

\def \y {{\bf y}}
\def \u {{\bf u}}

\def \n {{\bf n}}
\def \r {{\bf r}}
\def \t {{\bf t}}
\def \z {{\bf z}}
\def \e {{\bf e}}
\def \c {{\bf x}}

\def \ahat {\widehat{\a}}
\def \bhat {\widehat{\b}}

\newcommand{\abs}[1]{\left\vert#1\right\vert}
\newcommand{\norm}[1]{\left\|#1\right\|}

\newtheorem{THEO}{Theorem}
\newtheorem{LEMM}{Lemma}

\newtheorem{DEFI}{Definition}

\newtheorem{COROLLARY}{Corollary}

\newcommand{\bigo}[1]{\mathcal{O}\left(#1\right)}

\newcommand{\qed}{{\unskip\nobreak\hfil\penalty50\hskip2em\vadjust{}
          \nobreak\hfil$\Box$\parfillskip=0pt\finalhyphendemerits=0\par}}

\newcommand{\qq}{\vspace{-0mm}}  
\begin{document}

\title{{\bf Signal Recovery on Incoherent Manifolds}}
\author{{\em Chinmay Hegde and Richard G. Baraniuk} $\:$\thanks{Email:
\{chinmay, richb\}@rice.edu. Web: dsp.rice.edu/cs.
This work was supported by the grants NSF CCF-0431150, CCF-0926127, and CCF-1117939;
DARPA/ONR N66001-11-C-4092 and N66001-11-1-4090;
ONR N00014-08-1-1112, N00014-10-1-0989, and N00014-11-1-0714; 
AFOSR FA9550-09-1-0432; 
ARO MURI W911NF-07-1-0185 and W911NF-09-1-0383; 
and the Texas Instruments Leadership University Program. Thanks to Christoph Studer
for valuable comments on an early draft of the manuscript.}
\\
Department of Electrical and Computer Engineering \\
Rice University \vspace{-0.0in}}
\date{January 2012; Revised June 2012}

\maketitle

\begin{abstract}

Suppose that we observe noisy linear measurements of an unknown signal that can be modeled as the sum of two component signals, each of which arises from a nonlinear sub-manifold of a high-dimensional ambient space. We introduce Successive Projections onto INcoherent manifolds (SPIN), a first-order projected gradient method to recover the signal components. Despite the nonconvex nature of the recovery problem and the possibility of underdetermined measurements,  SPIN provably recovers the signal components, provided that the signal manifolds are incoherent and that the measurement operator satisfies a certain restricted isometry property. SPIN significantly extends the scope of current recovery models and algorithms for low-dimensional linear inverse problems and matches (or exceeds) the current state of the art in terms of performance.

\end{abstract}

\section{Introduction}
\label{sec:intro}

\subsection{Signal recovery from linear measurements}

Estimation of an unknown signal from linear observations
is a core problem in signal processing, statistics, and information theory. 
Particular energy has
been invested in problem instances where the available information is
{\em limited and noisy} and where the signals of interest possess a
{\em low-dimensional} geometric structure. 
Indeed, focused efforts on certain instances
of the linear inverse problem framework have spawned entire research
subfields, encompassing both theoretical and
algorithmic advances. Examples include signal separation and
morphological component analysis~\cite{DonohoBP,mca}; 
sparse approximation and compressive sensing~\cite{Tropp04,CandesCS,DonohoCS}; 
affine rank minimization~\cite{fazel2008compressed}; and robust
principal component analysis~\cite{candes2009robust,chandrasekaran2009sparse}. 

In this work, we study a very general version of the linear
inverse problem. Suppose that the signal of
interest $\x^*$ can be written as the sum of two constituent signals
$\a^* \in \A$ and $\b^* \in \B$, where $\A, \B$ are 
nonlinear, possibly non-differentiable, sub-manifolds of the signal space $\real^N$.
Suppose that we are given access to noisy linear measurements of $\x^*$:
\begin{equation}
\z = \bPhi(\a^* + \b^*) + \e,
\label{eq:meas}
\end{equation}
where $\bPhi \in \real^{M \times N}$ is the measurement matrix. Our objective is to recover the
pair of signals $(\a^*, \b^*)$, and thus also $\x^*$, from $\z$.
At the outset, numerous obstacles arise while trying to solve
(\ref{eq:meas}), some of which appear to be insurmountable:

\begin{enumerate}
\item (Identifiability I) Consider even the simplest case, where the measurements are noiseless
  and the measurement operator is the identity, i.e., we observe $\c \in \real^N$ such that
\begin{equation}
\c = \a^* + \b^*,
\label{eq:full}
\end{equation}
where $\a^* \in \A, \b^* \in \B$. This expression for $\c$ contains $2N$ unknowns but only $N$ observations and
hence is fundamentally ill-posed. Unless we make additional
assumptions on the geometric structure of the
component manifolds $\A$ and $\B$, a unique decomposition of $\c$ into
its constituent signals $(\a^*, \b^*)$ may not exist. 

\item (Identifiability II) To complicate matters, in more general situations the linear
  operator $\bPhi$ in (\ref{eq:meas}) might have fewer rows that
  columns, so that $M < N$. Thus, $\bPhi$ possesses a nontrivial nullspace. Indeed, we
are particularly interested in cases where $M \ll N$, in which case the nullspace
of $\Phi$ is extremely large relative to the ambient space. This
further obscures the issue of identifiability of the ordered pair
$(\a^*,\b^*)$, given the available observations $\z$.

\item (Nonconvexity) Even if the above two identifiability issues were resolved, the manifolds
$\A, \B$ might be extremely nonconvex, or even
non-differentiable. Thus, classical numerical methods, such as Newton's
method or steepest descent, cannot be successfully applied; neither can the litany of
convex optimization methods that have been specially designed for
linear inverse problems with certain types of signal priors~\cite{DonohoBP,fazel2008compressed}.

\end{enumerate}
In this paper, we propose a simple method to
recover the component signals $(\a^*, \b^*)$ from $\z$ in
(\ref{eq:meas}). 
We dub our method {\em Successive Projections onto
INcoherent manifolds} (SPIN) (see Algorithm~\ref{alg:spam}) 
Despite the highly nonconvex nature of the problem and the
possibility of underdetermined measurements, SPIN {\em provably} recovers the signal components
$(\a^*,\b^*)$. For this to hold true, we will require that (i) the
signal manifolds $\A,\B$ are {\em incoherent} in the sense that the
secants of $\A$ are almost orthogonal to the secants of $\B$;  
and (ii) the measurement operator $\bPhi$ satisfies a certain {\em
  restricted isometry property} (RIP) on the secants of the direct sum
manifold $\C = \A \oplus \B$. We will formally define these conditions
in Section~\ref{sec:geom}. We prove the following theoretical
statement below in Section~\ref{sec:main}.

\begin{THEO}[Signal recovery]
\label{thm:vanilla}
Let $\A, \B$ be incoherent manifolds in $\real^N$. Let
$\bPhi$ be a measurement matrix that satisfies the RIP on the direct sum $\C = \A \oplus \B$.  
Suppose we observe linear measurements $\z = \bPhi(\a^* +
\b^*)$, where $\a^* \in \A$ and $\b^* \in \B$.  Then, given any precision
parameter $\nu > 0$, there exists a positive integer $T_{\nu}$ and an iterative
algorithm that outputs a sequence of iterates $(\a_k, \b_k) \in \A
\times B, k = 1,2,\ldots$ such that $\max\{
\norm{\a_k - \a^*}, \norm{\b_k - \b^*}\} \leq 1.5 \nu$ for all
$k > T_{\nu}$. 
\end{THEO}

Our proposed algorithm (SPIN) is iterative in nature. Each iteration consists of three steps: computation of
the gradient of the error function $\psi(\a,\b) = \frac{1}{2}\norm{\z
  - \bPhi(\a+\b)}^2$, forming signal proxies for $\a$ and $\b$, and
orthogonally projecting the proxies onto the manifolds $\A$ and $\B$. 
The projection operators onto the component manifolds play a crucial
role in algorithm stability and performance; some manifolds admit 
stable, efficient projection operators while others do not. 
We discuss this in detail in Section~\ref{sec:main}. 
Additionally, we demonstrate that SPIN is stable to measurement noise
(the quantity $\e$ in (\ref{eq:meas})) as well as numerical inaccuracies (such
as finite precision arithmetic). 

\subsection{Prior Work}
The core essence of our proposed approach has been extensively studied in a
number of different contexts. Methods such as Projected Landweber
iterations~\cite{Daubechies-Defrise-DeMol-04}, iterative hard thresholding
(IHT)~\cite{IHT}, and singular value projection (SVP)~\cite{meka} are all instances of the
same basic framework.   SPIN subsumes and generalizes these methods. 
In particular, SPIN is an iterative projected gradient method
with the same basic approach as two recent signal recovery algorithms ---
Gradient Descent with Sparsification (GraDeS)~\cite{grades}, and  
Manifold Iterative Pursuit (MIP)~\cite{venkatallerton}.
We generalize these approaches to situations where the signal of interest
is a linear mixture of signals arising from a pair of nonlinear
manifolds. Due to the particular structure of our setting, SPIN consists of {\em two} projection steps (instead of
one), and the analysis is more involved (see Section~\ref{sec:theory}). 
We also explore the interplay between the geometric structure of the
component manifolds, the linear measurement operator, and the
stability of the recovery algorithm.

SPIN exhibits a strong geometric convergence rate comparable to many state-of-the-art first-order
methods~\cite{IHT,meka}, despite the nonlinear and nonconvex nature
of the reconstruction problem.  We duly note that, for the case of certain special manifolds, 
sophisticated higher-order recovery methods with stronger stability guarantees
have been proposed (e.g., approximate message passing (AMP)~\cite{amp}
for sparse signal recovery and augmented Lagrangian multiplier (ALM)
methods for low-rank matrix recovery~\cite{lin2010augmented}); see also~\cite{tropp2012}.
However, an appealing feature of SPIN is its conceptual
simplicity plus its ability to generalize to mixtures of arbitrary
nonlinear manifolds, provided these manifolds satisfy certain geometric
properties, as detailed in Section~\ref{sec:geom}.

\subsection{Setup}

For the rest of the paper, we will adopt the convention that vector- and
matrix-valued quantities appear in boldface (e.g., $\x, \y, \bPhi,
\ldots$), while scalar-valued quantities appear in standard italics
(e.g., $\alpha, \beta, k, K, \ldots$). Unless otherwise
specified, we will assume that $\| \cdot \|$ represents the Euclidean
norm (or the 2-norm) in $\real^N$ and that $\langle \cdot, \cdot \rangle$
represents the Euclidean inner product.

We are interested in ensembles of signals that can be modeled as
{\em low-dimensional manifolds} belonging to the signal
space. Informally, manifold signal models are applicable when (i) a $K$-dimensional
parameter vector $\btheta$ can be identified that captures the information present in a
signal, and (ii) the signal $\x = f(\btheta) \in \real^N$ can be {\em locally} modeled as a
continuous, possibly nonlinear, function $f$ of the parameters $\theta$. In such a 
scenario, the signal ensemble can be denoted as a $K$-dimensional
manifold $\M \in \real^N$. 
In our framework, we do not assume that the
function $f$ is smooth, i.e., the manifold $\M$ need not be a
Riemannian manifold. Examples of signal manifolds, as defined within
our framework, include the set of all sparse signals; the algebraic variety of all low-rank
matrices~\cite{chandrasekaran2009sparse}; and signal/image articulation manifolds
(see Section~\ref{subsec:am}). For an excellent introduction to
manifold-based signal models, refer to~\cite{wakinphd}.

\section{Geometric Assumptions}
\label{sec:geom}

The analysis and proof of accuracy of SPIN (Algorithm~\ref{alg:spam}) involves
three core ingredients: (i)  a geometric notion of {\em
  manifold incoherence} that crystallizes the approximate
orthogonality between secants of submanifolds of $\real^N$; (ii) a
{\em restricted isometry} condition satisfied by the measurement
operator $\bPhi$ over the secants of a submanifold; and (iii) the
availability of {\em projection operators} that compute the orthogonal
projection of any point $\x \in \real^N$ onto a submanifold of
$\real^N$.

\subsection{Manifold incoherence}

In linear inverse problems such as sparse signal approximation and compressive
sensing, the assumption of incoherence between linear subspaces,
bases, or dictionary elements is common. We introduce a nonlinear
generalization of this concept.

\begin{DEFI} 
Given a manifold $\A \subset \real^N$, a {\em normalized secant}, or
simply, a {\em secant}, $\u
\in \real^N$ of $\A$ is a unit vector such that
$$
\u = \frac{\a - \a'}{\norm{\a - \a'}},~~~\a, \a' \in \A,~\a \neq \a'.
$$
The {\em secant manifold} $\S(\A)$ is the family of 
unit vectors $\u$ generated by all pairs $\a, \a'$ in $\A$.
\end{DEFI}

\begin{DEFI} 
Suppose $\A, \B$ are submanifolds of $\real^N$. Let
\begin{equation}
\sup_{\u \in \S(\A),~\u' \in \S(\B)} \abs{\langle \u, \u' \rangle} =
\epsilon ,
\label{eq:inch}
\end{equation}
where $\S(\A), \S(\B)$ are the secant manifolds of $\A, \B$
respectively. Then, $\A$ and $\B$ are called
$\epsilon$-{\em incoherent} manifolds.
\end{DEFI}

Informally, any point on the secant manifold $\S(\A)$ represents a 
direction that aligns with a difference vector of $\A$, while the
incoherence parameter $\epsilon$ controls the extent of ``perpendicularity'' between the manifolds $\A$ and $\B$. We
define $\epsilon$ in terms of a supremum over sets $\S(\A),
\S(\B)$. Therefore,  a small value of $\epsilon$ implies that {\em
  each} (normalized) secant of $\A$ is approximately orthogonal to {\em all} secants of $\B$. 
By definition, the quantity $\epsilon$ is always non-negative; further,
$\epsilon \leq 1$, due to the Cauchy-Schwartz inequality.

We prove that any signal $\c$
belonging to the direct sum $\A \oplus \B$ can be {\em uniquely}
decomposed into its constituent signals when the upper bound on
$\epsilon$ holds with strict inequality.
\begin{LEMM}[Uniqueness]
Suppose that $\A, \B$ are $\epsilon$-incoherent with $0 < \epsilon < 1$. Consider $\c = \a + \b = \a' + \b'$, where $\a,
\a' \in \A$ and $\b, \b' \in \B$. Then, $\a = \a',
\b = \b'$.
\label{lem:uniq}
\end{LEMM}
{\em Proof.} It is clear that $\norm{\a + \b - (\a' +
  \b')}^2 = 0$, i.e., 
$$
\norm{\a-\a'}^2 + \norm{\b-\b'}^2 = -2 \langle \a-\a',\b-\b' \rangle
\leq 2 \abs{\langle \a-\a',\b-\b' \rangle}.
$$
However, due to the manifold incoherence assumption, the (unnormalized) secants $\a-\a'$, $\b-\b'$
obey the relation:
\begin{equation}
\abs{\langle \a-\a',\b-\b' \rangle} \leq \epsilon \norm{\a - \a'} \norm{\b -
  \b'}  \leq \frac{1}{2} \epsilon (\norm{\a-\a'}^2 +
\norm{\b-\b'}^2) , \label{eq:amgm}
\end{equation}
where the last inequality follows from the relation between arithmetic
and geometric
means (henceforth referred to as the {\em AM-GM inequality}). Therefore, we have that
$$
\norm{\a-\a'}^2 + \norm{\b-\b'}^2 \leq \epsilon  \left(\norm{\a-\a'}^2 + \norm{\b-\b'}^2\right),
$$
for $\epsilon < 1$, which is impossible unless $\a = \a', \b =
\b'$. \qed 

We can also prove the following relation between secants and direct sums
of signals lying on incoherent manifolds.
\begin{LEMM}
\label{lem:sec}
Suppose that $\A, \B$ are $\epsilon$-incoherent with $0 < \epsilon <
1$. Consider $\c_1 = \a_1 + \b_1, \c_2 = \a_2 + \b_2,$ where $\a_1,
\a_2 \in \A$ and $\b_1, \b_2 \in \B$. Then
$$
\abs{\langle \a_1 - \a_2, \b_1 - \b_2 \rangle} \leq
\frac{\epsilon}{2(1-\epsilon)}\norm{\c_1 - \c_2}^2 .
$$ 
\end{LEMM}
{\em Proof.} 
From (\ref{eq:amgm}), we have
\begin{eqnarray*}
\abs{\langle \a_1 - \a_2, \b_1 - \b_2 \rangle}&\leq&\frac{\epsilon}{2} (\norm{\a_1-\a_2}^2 +
\norm{\b_1-\b_2}^2) \\
&=&\frac{\epsilon}{2}\norm{\a_1+\b_1-\a_2-\b_2}^2 - \epsilon \langle
\a_1 -\a_2, \b_1 - \b_2 \rangle \\
&\leq& \frac{\epsilon}{2} \norm{\c_1-\c_2}^2 + \epsilon \abs{\langle
  \a_1 - \a_2, \b_1 - \b_2 \rangle} .
\end{eqnarray*}
Rearranging terms, we obtain the desired result. \qed

\subsection{Restricted isometry}

Next, we address the situation where the measurement operator $\bPhi
\in \real^{M \times N}$ contains a nontrivial nullspace, i.e., when $M < N$.
We will require that $\bPhi$ satisfies a {\em restricted isometry}
criterion on the secants of the direct sum manifold $\C = \A \oplus \B$.
\begin{DEFI}
Let $\C$ be a submanifold of $\real^N$. Then,  the matrix $\bPhi \in
\real^{M \times N}$ satisfies the
restricted isometry property (RIP) on $\C$ with constant $\delta \in
[0,1)$, if for every normalized secant $\u$ belonging to the secant
manifold $\S(\C)$, we have that
\begin{equation}
1 - \delta \leq \norm{\Phi \u}^2 \leq 1 + \delta .
\label{eq:rip}
\end{equation}
\end{DEFI}

The notion of restricted isometry (and its generalizations) is an important component in the
analysis of many algorithms in sparse approximation, compressive
sensing, and low-rank matrix
recovery~\cite{CandesCS,fazel2008compressed}. While the RIP has
traditionally been studied in the context of sparse signal models, (\ref{eq:rip})
generalizes this notion to {\em arbitrary} nonlinear
manifolds.  The restricted isometry condition is of particular interest when the
range space of the matrix $\bPhi$ is
low-dimensional.  A key result~\cite{BaraniukWakin06}
states that, under certain upper bounds on the curvature of the manifold
$\C$, there exist probabilistic
constructions of matrices $\bPhi$ that satisfy the RIP on $\C$ such that the number of rows of
$\bPhi$ is proportional to the intrinsic dimension of $\C$, rather
than the ambient dimension $N$ of the signal space.
We will discuss this further in Section~\ref{sec:app}.


\subsection{Projections onto manifolds}
 
Given an arbitrary nonlinear manifold $\A \in \real^N$, we define the 
operator $\P_\A(\cdot) : \real^N \mapsto \A$ as the Euclidean projection operator onto $\A$:
\begin{equation}
\P_\A(\x) = \arg \min_{\x' \in \A} \norm{\x' - \x}^2 .
\label{eq:proj}
\end{equation}
Informally, given an arbitrary vector $\x \in \real^N$, the operator
$\P_\A(x)$ returns the point on the manifold $\A$ that is ``closest''
to $\x$, where closeness is measured in terms of the Euclidean norm. 
Observe that for arbitrary nonconvex manifolds $\A$, the
above minimization problem (\ref{eq:proj}) may not yield a unique
optimum. Technically, 
therefore, $\P_\A(\x)$ is a set-valued operator. For ease of
exposition, $\P_\A(\x)$ will henceforth refer to {\em
  any} arbitrarily chosen element of the set of signals that
minimize the $\ell_2$-error in (\ref{eq:proj}).

The projection operator $\P_\A(\cdot)$ plays a crucial role in the development of our proposed
signal recovery algorithm in Section~\ref{sec:main}. Note that in a
number of applications, $\P_\A(\cdot)$ may be quite difficult to
compute exactly. The reasons for this might be intrinsic to the application
(such as the nonconvex, non-differentiable
structure of $\A$), or might be due to extrinsic constraints (such as
finite-precision arithmetic).
Therefore, following the lead of~\cite{venkatallerton}, we also define a $\gamma$-approximate
projection operator onto $\A$:
\begin{equation}
\x' = \P_\A^\gamma(\x) \implies \x' \in \A,~\textrm{and}~\norm{\x' - \x}^2 \leq \norm{\P_\A(\x)
- \x}^2 + \gamma,
\label{eq:projapprox}
\end{equation}
so that $ \P_\A^\gamma(\x)$ yields a vector $\x' \in \A$ that
approximately minimizes the squared distance from $\x$ to $\A$. Again, $\P_\A^\gamma(\x)$ need not be uniquely
defined for a particular input signal $\x$.

Certain specific instances of nonconvex manifolds do admit efficient exact projection
operators. For example, consider the space of all $K$-sparse signals
of length $N$; this can be viewed as the union of the ${N\choose K}$
canonical subspaces in $\real^N$ or, alternately, a $K$-dimensional
submanifold of $\real^N$. Then, the projection of an arbitrary vector
$\x \in \real^N$ onto this manifold is merely the best $K$-sparse
approximation to $\x$, which can be very efficiently computed via
simple thresholding. 
We discuss additional examples in Section~\ref{sec:app}.

\section{The SPIN Algorithm}
\label{sec:main}


We now describe an algorithm to solve the linear inverse
problem (\ref{eq:meas}). Our proposed algorithm, {\em Successive
 Projections onto INcoherent manifolds} (SPIN), can be viewed as a generalization of several
first-order methods for signal recovery for a variety of different
models~\cite{IHT,meka,venkatallerton}. 
SPIN is described in pseudocode form in Algorithm~\ref{alg:spam}. 

The key innovation in SPIN is that we formulate {\em two}
proxy vectors for the signal components $\widetilde{\a}_k$ and
$\widetilde{\b}_k$ and project these onto the corresponding manifolds $\A$ and $\B$. 
\begin{algorithm*}[!t]
\caption{Successive Projections onto INcoherent manifolds (SPIN) \label{alg:spam}}
\begin{algorithmic}
\STATE Inputs: Observation matrix $\bPhi$, measurements $\z$,
projection operators $\P_\A(\cdot), \P_\B(\cdot)$,
\STATE ~~~~~~~~~~~number of iterations $T$, step size $\eta$
\STATE Outputs: Estimated signal components $\ahat \in \A,\bhat \in \B$
\STATE Initialize: $\a_0= \mathbf{0}$, $\b_0 = \mathbf{0}$, $\r = \z$, $k = 0$
\WHILE{$k \leq T$}
\STATE $\g_k \leftarrow \eta \bPhi\trans \r$\qquad \qquad \qquad \qquad
\qquad \qquad \qquad~~~~~\COMMENT{form gradient}
\STATE $\widetilde{\a}_k \leftarrow \a_k + \g_k$,~$\widetilde{\b}_k \leftarrow \b_k
+ \g_k$ ~~~~\qquad \qquad \qquad \COMMENT{form signal proxies}
\STATE $\a_{k+1} \leftarrow \P_\A(\widetilde{\a}_k)$,~~$\b_{k+1}
\leftarrow \P_\B(\widetilde{\b}_k )$~~~~~~\qquad \qquad \COMMENT{apply
projection operators}
\STATE $\r \leftarrow \z - \bPhi(\a_{k+1} + \b_{k+1})$\qquad \qquad
\qquad \qquad \qquad\COMMENT{update residual}
\STATE $k \leftarrow k+1$
\ENDWHILE
\STATE return $(\ahat,\bhat) \leftarrow (\a_T, \b_T)$
\end{algorithmic}
\end{algorithm*}
We demonstrate that SPIN possesses strong {\em uniform recovery}
guarantees comparable to existing state-of-the-art algorithms for sparse
approximation and compressive sensing, while encompassing a very broad
range of nonlinear signal models.  The following theoretical result describes the performance of SPIN for
signal recovery.

\begin{THEO}[Main result]
\label{thm:recovery}
Suppose $\A, \B$ are $\epsilon$-incoherent manifolds in $\real^N$. Let
$\bPhi$ be a measurement matrix with restricted isometry constant
$\delta$ over the direct sum manifold $\C = \A \oplus \B$.  
Suppose we observe noisy linear measurements $\z = \bPhi(\a^* + \b^*)
+ \e$, where $\a^* \in \A$ and $\b^* \in \B$. 
 If  
\begin{equation}
0 \leq \delta < \frac{1 -
11\epsilon}{3 + 7\epsilon}, 
\label{eq:cond}
\end{equation}
then SPIN (Algorithm~\ref{alg:spam}) with
step size $\eta = 1/(1 + \delta)$ with exact projections $\P_\A,
\P_\B$ outputs $\a_T \in \A$ and $\b_T \in \B$, such that
$\norm{\z - \bPhi(\a_T + \b_T)}^2 \leq \beta \norm{\e}^2 + \nu$ in no
more than $T =
\lceil \frac{1}{\log(1/\alpha)} \log{\frac{\norm{\z}^2}{2\nu}} \rceil$ iterations for any $\nu > 0$.  
\end{THEO}

Here, $\alpha < 1$ and $\beta$ are moderately-sized positive constants that
depend only on $\delta$ and $\epsilon$; we derive explicit expressions
for $\alpha$ and $\beta$ in
Section~\ref{sec:theory}. For
example, when $\epsilon = 0.05,~\delta = 0.5$, we obtain $\alpha
\approx 0.812,~\beta \approx 5.404$. 

For the special case when there is no measurement noise (i.e.,
$\e=0$), Theorem~\ref{thm:recovery} states that, after a finite number of
iterations, SPIN outputs signal component estimates
$(\ahat,\bhat)$ such that $\norm{\z - \bPhi(\ahat + \bhat)} < \nu$ for
any desired precision parameter $\nu$. From the restricted isometry
assumption on $\bPhi$ and Lemma~\ref{lem:uniq}, we immediately obtain Theorem~\ref{thm:vanilla}.
Since we can set $\nu$ to an arbitrarily small value, we have that the
SPIN estimate $(\ahat,\bhat)$ converges to the true signal pair $(\a^*,\b^*)$. 
Exact convergence of the
algorithm might potentially take a very large number of iterations,
but convergence to any desired positive precision constant $\beta$
takes only a finite
number of iterations. For the rest of the paper, we will informally
denote signal ``recovery'' to imply convergence to a sufficiently fine precision.

SPIN assumes the availability of the exact
projection operators $\P_\A, \P_\B$. In certain cases, it might be
feasible to numerically compute only $\gamma$-approximate projections,
as in (\ref{eq:projapprox}). 
In this case, the bound on the norm of the error $\z - \bPhi(\a_T + \b_T)$ is
only guaranteed to be upper bounded by a positive multiple of the approximation
parameter $\gamma$.
The following theoretical guarantee (with a near-identical proof
mechanism as Theorem~\ref{thm:recovery}) captures this behavior.

\begin{THEO}[Approximate projections]
\label{thm:approx}
Under the same suppositions as Theorem~\ref{thm:recovery}, SPIN (Algorithm~\ref{alg:spam}) with
$\gamma$-approximate projections and step size $\eta = 1/(1 + \delta)$
outputs $\a_T \in \A$ and $\b_T \in B$ such that
$\norm{\z - \bPhi(\a_T + \b_T)}^2 \leq \beta \norm{\e}^2 +
\frac{1+\delta}{1-\alpha}\gamma + \nu$,  in no more than $T =
\lceil \frac{1}{\log(1/\alpha)} \log{\frac{\norm{\z}^2}{2\nu}} \rceil$ iterations.
\end{THEO}

We note some implications of Theorem~\ref{thm:recovery}.
 First, suppose that $\Phi$ is
the identity operator, i.e., we have full measurements of the signal
$\c^* = \a^* + \b^*$. Then, $\delta = 0$ and the lower bound on the
restricted isometry constant holds with equality. However, we still
require that $\epsilon < 1/11$ for guaranteed recovery using
SPIN. We will discuss this condition further in Section~\ref{sec:app}.

Second, suppose that the one of the component manifolds is the trivial
(zero) manifold; then, we have that $\epsilon = 0$. In this case, SPIN reduces to the Manifold Iterative Pursuit (MIP) algorithm for recovering signals from
a single manifold~\cite{venkatallerton}. Moreover, the condition on $\delta$ reduces to $0 \leq \delta < 1/3$, which
exactly matches the condition required for guaranteed recovery using MIP.

Lastly, the condition (\ref{eq:cond}) in Theorem~\ref{thm:recovery} automatically implies that $\epsilon <
1/11$. This represents a mild tightening of the condition on $\epsilon$
required for a unique decomposition (Lemma~\ref{lem:uniq}), even with
full measurements (i.e., when $\bPhi$ is the identity operator or,
more generally, when $\delta = 0$).

\section{Analysis}
\label{sec:theory}

The analysis of SPIN is based on the proof technique developed by~\cite{grades} for analyzing the
iterative hard thresholding (IHT) algorithm for sparse recovery and
further extended in~\cite{meka,venkatallerton}. For a given set of
measurements $\z$ obeying (\ref{eq:meas}), define the error function $\psi
: \A \times B \rightarrow \real$ as
$$
\psi(\a, \b) = \frac{1}{2}\norm{\z - \bPhi(\a + \b)}^2 . 
$$
It is clear that $\psi(\a^*, \b^*) = \frac{1}{2}\norm{\e}^2$. The
following lemma bounds the error of the estimated signals output by
SPIN at the $(k+1)$-st iteration in terms of the error incurred at the $k$-th
iteration, and the norm of the measurement error.

\begin{LEMM}
\label{lem:iter}
Define $(\a_k, \b_k)$ as the intermediate estimates obtained by SPIN
at the $k$-th iteration. Let $\delta, \epsilon$ be as defined in
Theorem~\ref{thm:recovery}. 
 Then,
\begin{equation}
\psi(\a_{k+1}, \b_{k+1}) \leq \alpha \psi(\a_k,\b_k) + C
\norm{\e}^2, 
\label{eq:iter}
\end{equation}
where
$$
\alpha = \frac{\frac{2\delta}{1 - \delta} +
  6\frac{1+\delta}{1-\delta}\frac{\epsilon}{1-\epsilon}}{1 -
  4\frac{1+\delta}{1-\delta}\frac{\epsilon}{1-\epsilon}} , ~~
C = \frac{\frac{1}{2} +
  5\frac{1+\delta}{1-\delta}\frac{\epsilon}{1-\epsilon}}{1 -
  4\frac{1+\delta}{1-\delta}\frac{\epsilon}{1-\epsilon}} .
$$
\end{LEMM}
{\em Proof.} Fix a current estimate of the signal components
$(\a_k,\b_k)$ at iteration $k$. Then, for any other pair of signals $(\a,\b) \in \A
\times \B$, we have
\begin{eqnarray*}
\psi(\a,\b) - \psi(\a_k,\b_k) & = &\frac{1}{2} \left( \norm{\z - \bPhi (\a +
  \b) }^2 - \norm{\z - \bPhi (\a_k + \b_k)}^2 \right) \\
 & = &\frac{1}{2} \left( \norm{\z - \bPhi (\a +
  \b) - \bPhi (\a_k + \b_k) + \bPhi (\a_k + \b_k)  }^2 - \norm{\z -
  \bPhi (\a_k + \b_k)}^2 \right) \\
& = & \frac{1}{2} \left( \norm{\z - \bPhi (\a_k + \b_k)}^2  + \norm{\bPhi
    (\a_k + \b_k) - \bPhi (\a + \b)}^2  - \norm{\z - \bPhi (\a_k + \b_k)}^2 \right) \\
& &~~+~\langle \z - \bPhi (\a_k + \b_k), \bPhi (\a_k + \b_k) - \bPhi (\a +
\b) \rangle \\
&=& \frac{1}{2}\norm{\bPhi \c - \bPhi \c_k}^2 + \langle \z - \bPhi \c_k,
\bPhi \c_k - \bPhi \c \rangle , 
\end{eqnarray*}
where $\c_k \triangleq \a_k + \b_k,~\c \triangleq \a + \b$. 
Since $\bPhi$ is a linear operator, we can
take the adjoint within the inner product to obtain
\begin{eqnarray}
\psi(\a,\b) - \psi(\a_k,\b_k) &=& \frac{1}{2}\norm{\bPhi \c - \bPhi
  \c_k}^2 +  \langle \bPhi\trans ( \z - \bPhi \c_k ),
 \c_k -  \c \rangle \label{eq:psisq0} \\
&\leq& \frac{1}{2} (1 + \delta) \norm{\c - \c_k}^2 +  \langle \bPhi\trans ( \z - \bPhi \c_k ),
 \c_k -  \c \rangle \label{eq:psisq}  .
\end{eqnarray}
The last inequality occurs due to the RIP of $\bPhi$
applied to the secant vector $\c -\c_k \in \S(\C)$. To the right hand
side of (\ref{eq:psisq}), we
further add and subtract $\frac{1}{2(1+\delta)}
\norm{\bPhi\trans(\z - \bPhi \c_k)}^2 $ to complete the square:
$$
\psi(\a,\b) - \psi(\a_k,\b_k) \leq \frac{1}{2} (1 + \delta) \norm{\c -
  \c_k - \frac{1}{1+\delta}{\bPhi\trans(\z - \bPhi \c_k)} }^2 -
\frac{1}{2(1+\delta)} \norm{\bPhi\trans(\z - \bPhi \c_k )}^2 .
$$
Define $\g_k \triangleq \frac{1}{1+\delta}\bPhi\trans(\z - \bPhi(\a_k + \b_k))$. Then,
\begin{equation}
\psi(\a,\b) - \psi(\a_k,\b_k)  \leq \frac{1}{2} (1 + \delta) \left( \norm{ \a
  + \b - (\a_k + \b_k + \g_k) }^2 - \norm{\g_k}^2 \right) .
\label{eq:psi}
\end{equation}
Next, define the function $\zeta$ 
on $\A \times \B$ as $\zeta(\a,\b) \triangleq \norm{ \a
  + \b - (\a_k + \b_k + \g_k) }^2$. Then, we have
\begin{eqnarray*}
\zeta(\a_{k+1},\b_{k+1}) &=& \norm{ \a_{k+1} - (\a_k +
    \g_k) + \b_{k+1} - (\b_k + \g_k) + \g_k}^2 \\
&=& \norm{\a_{k+1} - (\a_k + \g_k)}^2
  + \norm{\b_{k+1} - (\b_k + \g_k)}^2 + \norm{\g_k}^2 \\
&& + 2 \langle \a_{k+1} - (\a_k
  + \g_k), \b_{k+1} - (\b_k + \g_k) \rangle + 2 \langle \g_k, \a_{k+1} + \b_{k+1} -
  (\a_k + \b_k + 2 \g_k) \rangle.
\end{eqnarray*}
But, as specified in Algorithm~\ref{alg:spam}, $\a_{k+1} = \P_\A(\a_k + \g_k)$, and hence
$\norm{\a_{k+1} - (\a_k + \g_k)} \leq \norm{\a - (\a_k + \g_k)}$ for
any $\a \in \A$. An analogous relation can be formed between $\b_{k+1}$ and
$\b^*$. Hence, we have
\begin{eqnarray*}
\norm{\a_{k+1} - (\a_k + \g_k)} &\leq&\norm{\a^* - (\a_k + \g_k)}~~\textrm{and} \\
\norm{\b_{k+1} - (\b_k + \g_k)} &\leq&\norm{\b^* - (\b_k + \g_k)} .
\end{eqnarray*}
Substituting for $(\a_{k+1}, \b_{k+1})$, we obtain
\begin{eqnarray*}
\zeta(\a_{k+1}, \b_{k+1}) &\leq&  \norm{\a^* - (\a_k + \g_k)}^2
  + \norm{\b^* - (\b_k + \g_k)}^2 + \norm{\g_k}^2  \\
&& +~2 \langle \a_{k+1} - (\a_k
  + \g_k), \b_{k+1} - (\b_k + \g_k) \rangle + 2 \langle \g_k, \a_{k+1} + \b_{k+1} -
  (\a_k + \b_k + 2 \g_k) \rangle \\
&=& \norm{\a^* - (\a_k + \g_k)}^2
  + \norm{\b^* - (\b_k + \g_k)}^2 + \norm{\g_k}^2  \\
&& +~2 \langle \a^* - (\a_k
  + \g_k), \b^* - (\b_k + \g_k) \rangle + 2 \langle \g_k, \a^* + \b^* -
  (\a_k + \b_k + 2 \g_k) \rangle  \\
&&  +~ 2 \langle \a_{k+1} - (\a_k
  + \g_k), \b_{k+1} - (\b_k + \g_k) \rangle - 
2 \langle \a^* - (\a_k
  + \g_k), \b^* - (\b_k + \g_k) \rangle \\
&& + ~ 2 \langle \g_k, \a_{k+1} + \b_{k+1} -
  (\a^* + \b^*) \rangle.
\end{eqnarray*}
Completing the squares, we have:
\begin{eqnarray}
\zeta(\a_{k+1}, \b_{k+1}) &\leq& \norm{\a^* + \b^* - (\a_k + \b_k +
  \g_k )}^2 + 2 \langle \a_{k+1}-\a_k, \b_{k+1}-\b_k \rangle -2
\langle \a^*-\a_k, \b^*-\b_k \rangle \nonumber \\
&& +~ 2 \langle \g_k, -\a_{k+1} + \a_k - \b_{k+1} + \b_k + \a^* - \a_k +
\b^* - \b_k  + \a_{k+1} + \b_{k+1} - (\a^* + \b^*)   \rangle . \nonumber
\end{eqnarray}
The last term on the right hand side equals zero, and
so we obtain
$$
\zeta(\a_{k+1}, \b_{k+1}) \leq \zeta(\a^*, \b^*) +  2 \langle \a_{k+1}-\a_k, \b_{k+1}-\b_k \rangle -2
\langle \a^*-\a_k, \b^*-\b_k \rangle .
$$
Combining this inequality with (\ref{eq:psi}), we obtain the
series of inequalities
\begin{eqnarray}
\psi(\a_{k+1},\b_{k+1}) - \psi(\a_k,\b_k) & \leq & \frac{1}{2} (1 + \delta) \left(
\zeta(\a_{k+1}, \b_{k+1}) - \norm{\g_k}^2 \right) \nonumber \\
& \leq & \overbrace{\frac{1}{2} (1 + \delta) \left(\zeta(\a^*, \b^*) -
  \norm{\g_k}^2 \right)}^{\mathbb{T}_1}  \nonumber\\
&& + \overbrace{(1+\delta) \left( \langle \a_{k+1}-\a_k, \b_{k+1}-\b_k \rangle -
\langle \a^*-\a_k, \b^*-\b_k \rangle \right)}^{\mathbb{T}_2} \label{eq:rhs} \\
&=& \mathbb{T}_1 + \mathbb{T}_2 \nonumber.
\end{eqnarray}
We can further bound the right hand side of (\ref{eq:rhs}) as follows. 
First, we expand $\mathbb{T}_1$ to obtain
\begin{eqnarray*}
\mathbb{T}_1  &=& \frac{1}{2} (1 + \delta) \left(
  \norm{\a^*+\b^*-(\a_k+\b_k+\g_k)}^2 - \norm{\g_k}^2 \right) \\
&=&\frac{1}{2} (1 + \delta) \left( \norm{\a^* + \b^* -(\a_k + \b_k)}^2 -
2 \langle \g_k,  \a^* + \b^* -(\a_k + \b_k) \rangle \right) \\
&=& \frac{1}{2} (1 + \delta) \norm{\c^* - \c_k}^2 - \langle
\bPhi\trans(\z - \bPhi \c_k), \c^* - \c_k \rangle .
\end{eqnarray*}
Again, $\c^* - \c_k$ is a secant on the direct sum manifold
$\C$. By the RIP property of $\bPhi$, we have
\begin{eqnarray}
\mathbb{T}_1 & \leq & \frac{1}{2}\frac{1+\delta}{1-\delta}\norm{\bPhi \c^* -
  \bPhi \c_k}^2 +  \langle
\bPhi\trans(\z - \bPhi \c_k), \c_k - \c^* \rangle \nonumber \\
& = & \frac{1}{2}\left(\frac{1- \delta}{1 - \delta} + \frac{2\delta}{1-\delta}\right)\norm{\bPhi \c^* -
  \bPhi \c_k}^2 +  \langle
\bPhi\trans(\z - \bPhi \c_k), \c_k - \c^* \rangle \nonumber \\
&=&\frac{1}{2} \norm{\bPhi \c^* - \bPhi \c_k}^2 +  \langle
\bPhi\trans(\z - \bPhi \c_k), \c_k - \c^* \rangle +
\frac{\delta}{1-\delta}\norm{\z-\Phi \c_k}^2. \nonumber 
\end{eqnarray}
By definition, we have that $\psi(\a_k,\b_k) = \frac{1}{2}\norm{\z-\Phi
  \c_k}^2$. Further, we can substitute $\a = \a^*, \b = \b^*$ in
(\ref{eq:psisq0}) to obtain
\[
\psi(\a^*, \b^*) - \psi(\a_k, \b_k) = \frac{1}{2} \norm{\bPhi \c^* - \bPhi \c_k}^2 +  \langle
\bPhi\trans(\z - \bPhi \c_k), \c_k - \c^* \rangle.
\]
Therefore, we have the relation
\begin{equation}
\mathbb{T}_1 \leq \psi(\a^*, \b^*) - \psi(\a_k, \b_k) + \frac{2\delta}{1-\delta}\psi(\a_k, \b_k). \label{eq:t1}
\end{equation}
The term $\mathbb{T}_2$ can be bounded using Lemma~\ref{lem:sec} as
follows. We have
\begin{eqnarray}
- \langle \a^*-\a_k, \b^*-\b_k \rangle & \leq &
|\langle \a^*-\a_k, \b^*-\b_k \rangle|, \nonumber\\
&\leq& \frac{\epsilon}{2(1-\epsilon)}\norm{\c^*-\c_k}^2. \label{eq:t11}
\end{eqnarray}
Further, we have
\begin{eqnarray}
\langle \a_{k+1}-\a_k, \b_{k+1}-\b_k \rangle & \leq & | \langle
\a_{k+1}-\a_k, \b_{k+1}-\b_k \rangle | 
\leq \frac{\epsilon}{2(1-\epsilon)}\norm{\c_{k+1} -\c_{k}}^2
\nonumber \\
& = & \frac{\epsilon}{2(1-\epsilon)}\norm{(\c_{k+1} -\c^*) - (\c_k -
  \c^*) }^2 \nonumber \\
& = & \frac{\epsilon}{2(1-\epsilon)} \left( \norm{\c_{k+1} -\c^*}^2 +
  \norm{\c_{k} -\c^*}^2 - 2 \langle \c_{k+1} -\c^*, \c_{k} -\c^*
  \rangle \right) \nonumber \\
& \leq & \frac{\epsilon}{2(1-\epsilon)} \left( \norm{\c_{k+1} -\c^*}^2 +
  \norm{\c_{k} -\c^*}^2 + 2 \left|\langle \c_{k+1} -\c^*, \c_{k} -\c^*
  \rangle \right| \right)\nonumber \\
& \leq & \frac{\epsilon}{2(1-\epsilon)} \left( \norm{\c_{k+1} -\c^*}^2 +
  \norm{\c_{k} -\c^*}^2 + 2\norm{\c_{k+1} -\c^*} \norm{\c_{k} -\c^*
  \rangle} \right)\nonumber \\
&\leq&\frac{\epsilon}{(1-\epsilon)}\left(\norm{\c_{k+1}-\c^*}^2+\norm{\c_{k}-\c^*}^2
\right), \label{eq:trick} 
\end{eqnarray}
where the last two inequalities follow by applying the Cauchy-Schwartz inequality and the AM-GM inequality.  
Combining (\ref{eq:t11}) and (\ref{eq:trick}), $\mathbb{T}_2$
can be bounded above as
\begin{eqnarray*}
\mathbb{T}_2 &\leq& (1+\delta)\frac{\epsilon}{1-\epsilon} \left(
  \frac{3}{2} \norm{\c^*-\c_k}^2 + \norm{\c^*-\c_{k+1}}^2 \right), \\
& \leq & \frac{1+\delta}{1-\delta} \frac{\epsilon}{1-\epsilon} \left(
  \frac{3}{2} \norm{\bPhi\c^*-\bPhi\c_k}^2 +
  \norm{\bPhi\c^*-\bPhi\c_{k+1}}^2 \right) .
\end{eqnarray*}
But,
\begin{eqnarray*}
\norm{\bPhi\c^*-\bPhi\c_k}^2 &=&\norm{\bPhi\c^*-\bPhi\c_k + \e - \e}^2 \\
&\leq& 2 \norm{\bPhi\c^*-\bPhi\c_k + \e}^2 + 2\norm{\e}^2 =
4\psi(\a_k,\b_k)+4\psi(\a^*,\b^*)
\end{eqnarray*}
via the same technique used to obtain (\ref{eq:trick}). Similarly,
\begin{eqnarray*}
\norm{\bPhi\c^*-\bPhi\c_{k+1}}^2 &\leq&
4\psi(\a_{k+1},\b_{k+1})+4\psi(\a^*,\b^*) .
\end{eqnarray*}
Hence, we obtain
\begin{eqnarray}
\mathbb{T}_2 &\leq& \frac{1+\delta}{1-\delta}\frac{\epsilon}{1-\epsilon} \left(
\frac{3}{2} \left(4 \psi(\a_k, \b_k) +  4 \psi(\a^*, \b^*) \right) + 4 \psi(\a_{k+1}, \b_{k+1}) +  4 \psi(\a^*, \b^*)
\right)\nonumber \\
& = &\frac{1+\delta}{1-\delta} \frac{2\epsilon}{1-\epsilon} \left(
  3 \psi(\a_k,\b_k) + 2 \psi(\a_{k+1},\b_{k+1}) + 5 \psi(\a^*,\b^*)
\right) . \label{eq:t2}
\end{eqnarray}
Combining (\ref{eq:rhs}), (\ref{eq:t1}), and (\ref{eq:t2}), we obtain
\begin{eqnarray*}
\psi(\a_{k+1},\b_{k+1}) & \leq & \psi(\a^*,\b^*) +
\frac{2\delta}{1-\delta}\psi(\a_k,\b_k)  \\
&& +~\frac{1+\delta}{1-\delta} \frac{2\epsilon}{1-\epsilon} \big(
  3 \psi(\a_k,\b_k) + 2\psi(\a_{k+1},\b_{k+1}) + 5 \psi(\a^*,\b^*)
  \big).
\end{eqnarray*}
Rearranging, we obtain Lemma~\ref{lem:iter}. \qed 

{\em Proof of Theorem~\ref{thm:recovery}.} Equation~\ref{eq:iter}
describes a linear recurrence relation for the sequence of positive real
numbers $\psi(\a_k, \b_k),~k=0,1,2,\ldots$ with leading coefficient
$\alpha$. By choice of initialization, $\psi(\a_0, \b_0)
= \frac{\norm{\z}^2}{2}$. Therefore, for all $k \in \mathbb{N}$, we have the relation
 \begin{eqnarray*}
\psi(\a_k, \b_k) & \leq & \alpha^k  \psi(\a_0, \b_0) + C \frac{1 -
  \alpha^k}{1-\alpha} \norm{e}^2 \\
& \leq & \alpha^k \psi(\a_0, \b_0) + \frac{C}{1-
  \alpha} \norm{\e}^2 .
\end{eqnarray*}
To ensure that the value of $\psi(\a_k,\b_k)$ does not diverge, the
leading coefficient $\alpha$ must be smaller than 1, i.e., 
\[
{\frac{2\delta}{1 - \delta} +
  6\frac{1+\delta}{1-\delta}\frac{\epsilon}{1-\epsilon}} < 1 -
  4\frac{1+\delta}{1-\delta}\frac{\epsilon}{1-\epsilon} .
\]
Rearranging, we obtain the upper bound on $\delta$ as in (\ref{eq:cond}): 
\[ 
\delta < \frac{1 -
11\epsilon}{3 + 7\epsilon}.
\] 
By choosing $\beta = \frac{C}{1-\alpha}$, and  $k \geq T$ such that $T = \lceil \frac{1}{\log(1/\alpha)}
\log{\frac{\norm{\z}^2}{2\nu}} \rceil$, the result follows. \qed

The proof mechanism of Theorem~\ref{thm:approx} follows a near-identical 
procedure as in Lemma~\ref{lem:iter} and we omit the details for
brevity. Also, we observe that Theorem~\ref{thm:recovery} represents merely a
sufficient condition for signal recovery; the constants in (\ref{eq:cond}) could likely be
improved, but we will not pursue that direction in this paper. 

\section{Applications}
\label{sec:app}

The two-manifold signal model described in this paper is applicable to a wide variety of
problems that have attracted considerable interest in the literature
over the last several years. We discuss a few representative
instances and show how SPIN can be utilized for efficient signal
recovery in each of these instances. We also present several 
numerical experiments that indicate the kind of gains that SPIN can offer in practice.

\subsection{Sparse representations in pairs of bases}
We revisit the classical problem of decomposing signals in an
overcomplete dictionary that is the union of a pair of
orthonormal bases and show how SPIN can be used to efficiently solve
this problem. Let $\bPsi, \bPsi'$ be orthonormal bases of $\real^N$. Let
$\A$ be the set of all $K_1$-sparse signals in $\real^N$ in
the basis expansion of $\bPsi$, and let $\B$ be the set of all $K_2$-
sparse signals in the basis expansion of $\bPsi'$. Then, $\A$ and $\B$
can be viewed as $K_1$- and $K_2$-dimensional submanifolds of
$\real^N$, respectively. Consider a signal $\c^*
= \a^* + \b^*$, where $\a^* \in \A, \b^* \in \B$ so that
\[
\a^* = \sum_{i = 1}^{K_1} a_i \bpsi_i, ~~\b^* = \sum_{i =
  1}^{K_2} b_i \bpsi'_i .
\]
The problem is to recover $(\a^*, \b^*)$ given $\c^*$.  This problem
has been studied in many different forms in the literature, and several
algorithms have been proposed in order to solve it efficiently~\cite{DonohoBP,eladgeneralized,Tropp04}.  
 See~\cite{studer} for an in-depth
study of the various state-of-the-art methods. All these methods assume a certain notion of incoherence between the
two bases, most commonly referred to as the {\em mutual coherence}
$\mu$, which is defined as
\begin{equation} 
\mu(\bPsi,\bPsi') \triangleq \max_{i, j} \abs{\langle \bpsi_i, \bpsi'_j
  \rangle} .
\label{mutualinc}
\end{equation}
It is clear that $\mu \leq 1$. It can be shown that the
mutual coherence also obeys the lower bound $\mu \geq 1/\sqrt{N}$ for any
pair of bases of $\real^N$. This lower bound is in fact tight;
equality is achieved, for example, when $\bPsi$ is the canonical basis
in $\real^N$ and $\bPsi'$ is the basis defining the Walsh-Hadamard
transform or the discrete Fourier basis~\cite{eladgeneralized,Tropp04}.

We establish the following simple relation between $\mu$ and the manifold incoherence between $\A$ and
$\B$. 
\begin{LEMM}
Let $\A$ be the set of all $K_1$-sparse signals in $\bPsi$, and $\B$ be the set of all $K_2$-sparse signals in $\bPsi'$. Let $\epsilon$ denote the manifold
incoherence between $\A$ and $\B$. Then,
\label{lem:basisinc}
\[
\epsilon \leq \mu(\bPsi,\bPsi') (K_1 + K_2).
\]
\end{LEMM}
{\em Proof. } 
The secant manifold $\S(\A)$ is equivalent to the set of
signals in $\real^N$ that are $2K_1$-sparse in $\A$; similarly,
$\S(\B)$ is equivalent to the set of $2K_2$-sparse signals in
$\B$. Therefore, if one considers unit norm vectors $\u \in \S(\A), \u' \in \S(\B)$, we
obtain
\begin{eqnarray}
| \langle \u, \u' \rangle| &=& \left| \left\langle \sum_{i=1}^{2K_1} a_i \lambda_i,
\sum_{j=1}^{2K_2} b_j \lambda_j' \right\rangle \right| = \left| \sum_{i}^{{2K_1}} \sum_{j}^{{2K_2}} \langle \lambda_i,
\lambda'_j \rangle a_i b_j \right|  \nonumber \\
&\leq& \mu \sum_{i}^{{2K_1}} \sum_{j}^{{2K_2}} {|a_i b_j|} =
\mu \left( \sum_{i=1}^{2K_1}  |a_i| \right) \left( \sum_{j=1}^{2K_2}
  |b_j| \right), \label{eq:mu}
\end{eqnarray}
where the last relation follows from the triangle inequality. 
We can further bound the right hand side of (\ref{eq:mu}). We have
\[
\sum_{i=1}^{2K_1}  |a_i| \leq \sqrt{2K_1} \sqrt{\sum_{i=1}^{2K_1}
  |a_i|^2} = \sqrt{2K_1} \| \u \| =  \sqrt{2K_1},
\]
since $\u$ is a unit vector. Similarly, $\sum_{i=1}^{2K_2}  |b_i| \leq
\sqrt{2K_2}$. Inserting these upper bounds in (\ref{eq:mu}), we have
\[
| \langle \u, \u' \rangle| \leq \mu \sqrt{2K_1} \sqrt{2K_2} \leq \mu
(K_1 + K_2) .
\]
The lemma follows by considering the supremum over all vectors
$\u \in \S(\A), \u' \in \S(\B)$. \qed

We show how SPIN can be used to solve the linear inverse problem of
recovering $(\a^*, \b^*)$ from $\c^*$. The restricted isometry assumption is not relevant in this
case, since we assume that we have full measurements of the signal; therefore
$\delta = 0$. An upper bound for the manifold
incoherence parameter $\epsilon$ is specified in
Lemma~\ref{lem:basisinc}. The (exact) projection operators $\P_\A, \P_\B$ can
be easily implemented; we simply perform a coefficient expansion
in the corresponding orthornormal basis and retain the coefficients of largest magnitude. Mixing these
ingredients together, we can guarantee that, given any signal $\c^*$,
SPIN will return the true components $\a^*, \b^*$. This guarantee is summarized in the following result.

\begin{COROLLARY}[SPIN for pairs of bases]
\label{corr:basis}
Let $\c^* = \a^* + \b^*$, where $\a^*$ is $K_1$-sparse in $\bPsi$ and
$\b^*$ is $K_2$-sparse in $\bPsi'$. Let $\mu$ denote the mutual
coherence between $\bPsi$ and $\bPsi'$. Then, SPIN
exactly recovers ($\a^*, \b^*$) from $\c^*$ provided 
\begin{equation}
K_1 + K_2 < \frac{1}{11\mu} \approx \frac{0.091}{\mu}.
\label{eq:pairs}
\end{equation}
\end{COROLLARY}
{\em Proof.} If (\ref{eq:pairs}) holds, then from Lemma~\ref{lem:basisinc} we know that the
manifold incoherence $\epsilon$ between $\A$ and $\B$ is smaller
than 1/11. But this is exactly the condition required for guaranteed
convergence of SPIN to the true signal components $(\a^*, \b^*)$. \qed

SPIN thus offers a conceptually simple method to separate
mixtures of signals that are sparse in incoherent bases. The only
condition required on the signals is that the total sparsity $K_1 +
K_2$ is upper-bounded by the quantity $0.09/\mu$. The best
known approach for this problem is an
$\ell_1$-minimization formulation that features a similar guarantee that is known to be tight~\cite{eladgeneralized,feuer}:
$$K_1 + K_2 < \frac{\sqrt{2}-0.5}{\mu} \approx \frac{0.914}{\mu}.$$
Therefore, SPIN yields a recovery guarantee that is off the best possible
method by a factor of 10. 
Once again, it is possible that the constant $1/11$
in Corollary~\ref{corr:basis} can be tightened by a more careful
analysis of SPIN specialized to the case when the component signal
manifolds correspond to a pair of incoherent bases, but we will not pursue
this direction here. It is also possible to generalize SPIN to the case
where the sparsifying dictionary comprises a union of more than two
orthonormal bases~\cite{grib}; see
Section~\ref{sec:conc} for a short discussion.

\subsection{Articulation manifolds}
\label{subsec:am}

Articulation manifolds provide a powerful, flexible conceptual tool for modeling
signals and image ensembles in a number of applications~\cite{grimes,iam}.
Consider an ensemble of signals $\M \subset \real^N$ that are generated by
varying $K$ parameters $\theta \in \Theta,~\Theta
\subset \real^K$. 
Then, we say that the signals trace out a nonlinear $K$-dimensional
articulation manifold in $\real^N$, where $\theta$ is called the
{\em articulation parameter vector}. Examples of articulation manifolds
include: acoustic chirp signals of varying frequencies (where $\theta$ represents
the chirp rate); images of a white disk translating on a black background
(where $\theta$ represents the planar location of the disk center);
and images of a solid object with variable pose (where $\theta$ represents
the six-dimensional pose parameters, three corresponding to spatial location and
three corresponding to orientation). 

We consider the class of {\em compact, smooth}, $K$-dimensional articulation
manifolds $\M \subset \real^N$. For such manifold classes, it is possible to
construct linear measurement operators $\bPhi$ that preserve the
pairwise secant geometry of $\M$. 
Specifically, it has been shown~\cite{BaraniukWakin06} 
that there exist {\em randomized} constructions of measurement operators $\bPhi
\in \real^{M \times N}$ that satisfy the RIP on the secants of $\M$
with constant $\delta$ and with probability
at least $\rho$, provided
\[
M = \bigo{K \frac{\log(C_\M N \delta^{-1}) \log(\rho^{-1})}{\delta^2} }
\]
for some constant $C$ that depends only on the smoothness and volume of the
manifold $\M$. Therefore, the dimension of the range space of $\bPhi$ is proportional
to the {\em number of degrees of freedom} $K$, but is only
logarithmic in the ambient dimension $N$. Moreover, given such a
measurement matrix $\bPhi$ with isometry constant
$\delta < 1/3$ and a projection operator $\P_\M(\cdot)$ onto $\M$, any signal
$\x \in \M$ can be reconstructed from its compressive measurements $\y
= \bPhi \x$ using Manifold Iterative Pursuit (MIP)~\cite{venkatallerton}.


\begin{figure*}[!t]
\centering
\begin{tabular}{ccc}
\centering
\includegraphics[width=0.23\textwidth]{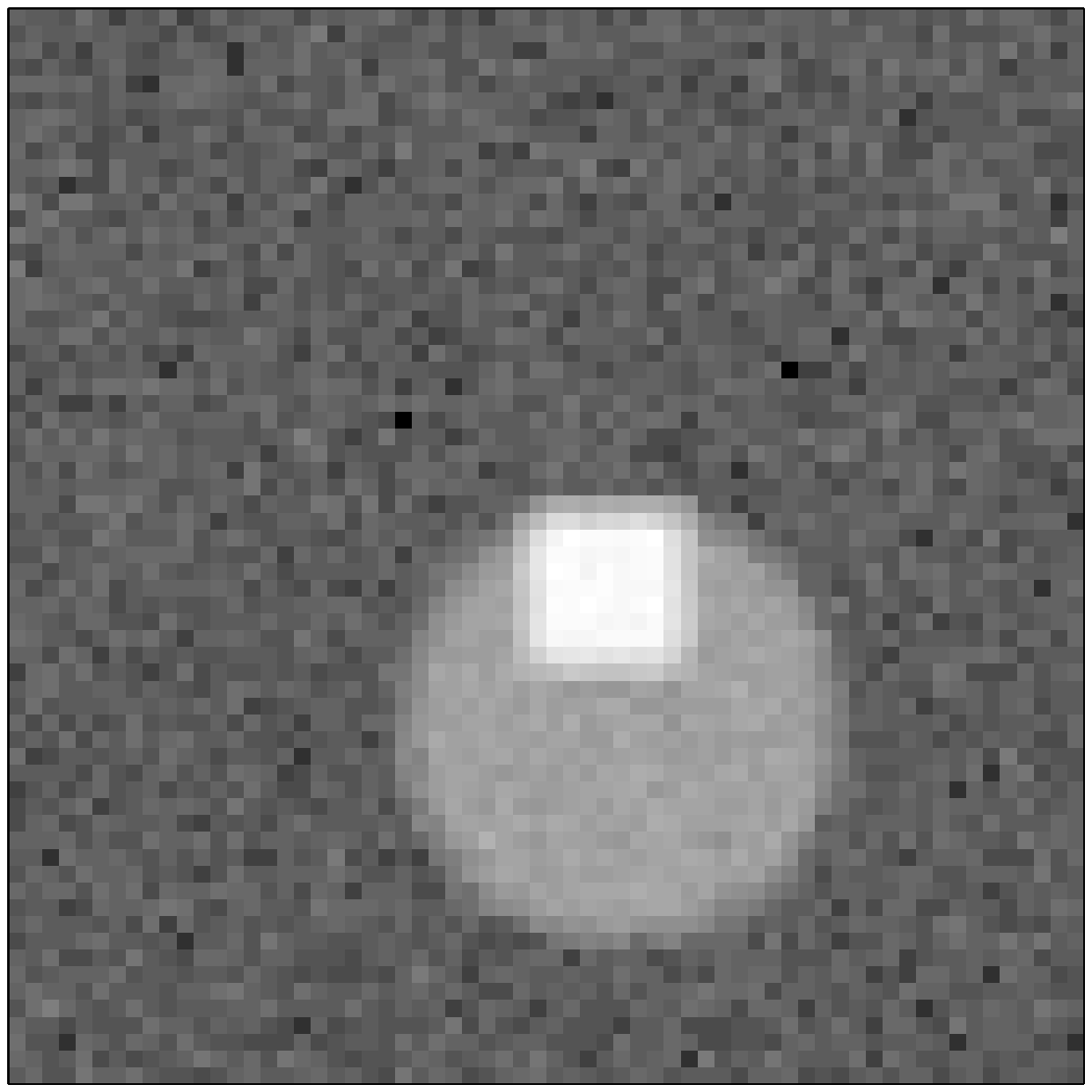} &
\includegraphics[width=0.23\textwidth]{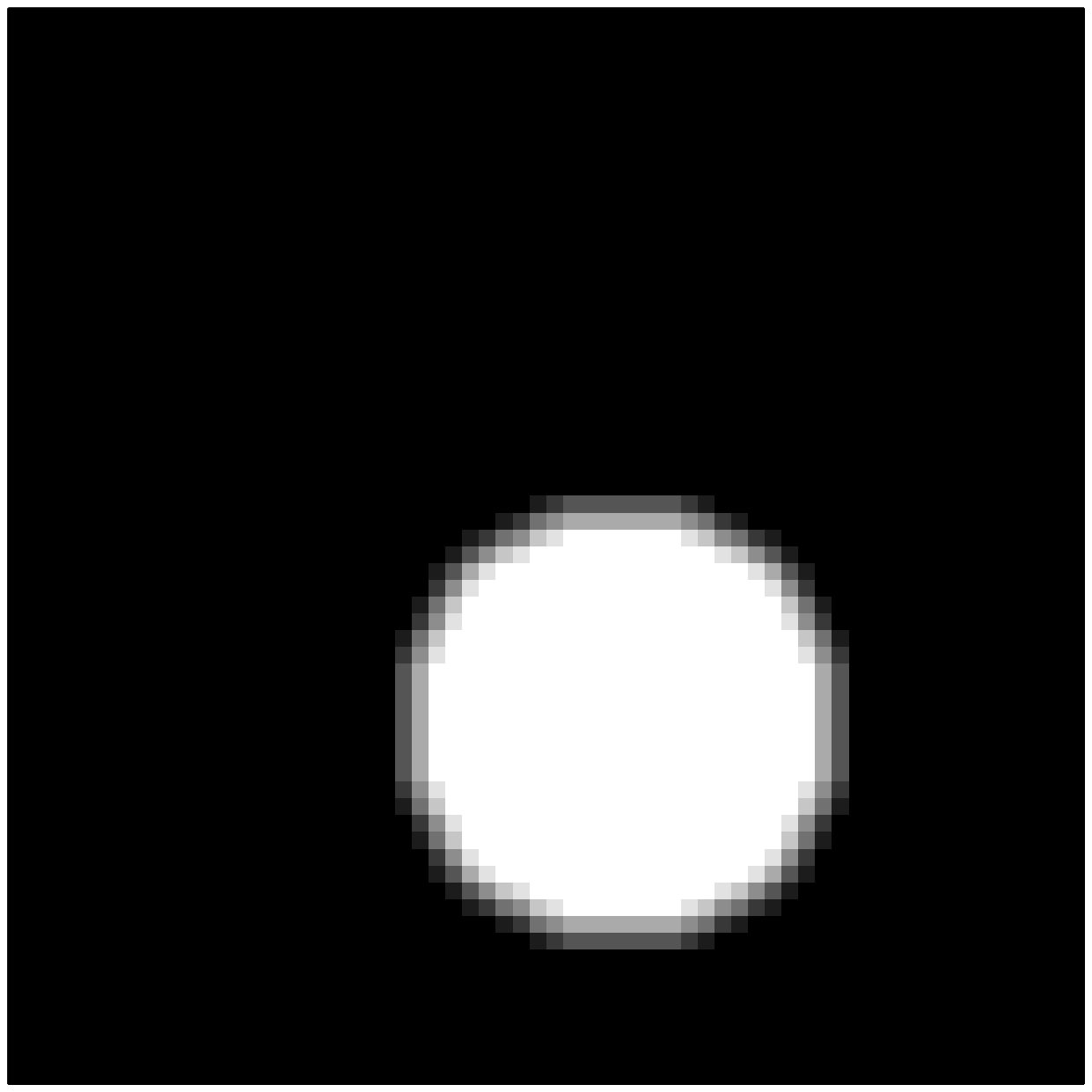} &
\includegraphics[width=0.23\textwidth]{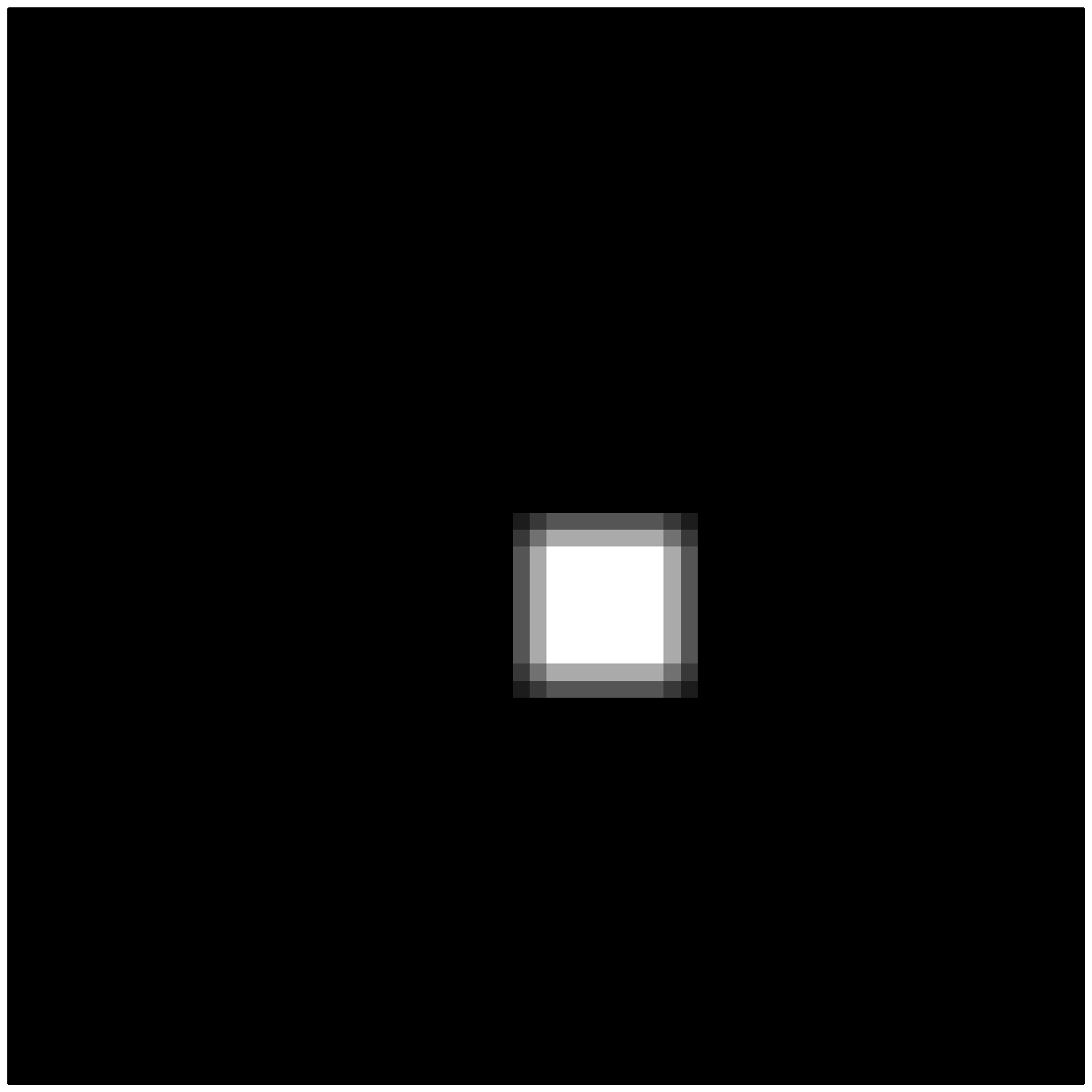} \\
{\small (a) Original image} & {\small (b) Disk component} &
{\small (c) Square component}
\end{tabular}
\caption{{\small\sl SPIN recovery of a noisy $64 \times 64$ image from compressive
    measurements. The clean image consists of the linear superposition of a
    disk and a square of fixed pre-specified sizes, but the locations of the centers of the disk and the square are
    unknown. Additive Gaussian noise (SNR = 14dB) is added to the
    image prior to measurement.
Signal length $N = 64\times64=4096$, number of compressive
measurements $M = 50.$ (a)~Original noisy image. (b)~Reconstructed disk. (c)~Reconstructed square. SPIN perfectly
reconstructs both components from just M/N = 1.2\% measurements.}
\label{fig:circsq}
}
\end{figure*}

We generalize this setting to the case where the unknown signal of interest
arises as a mixture of signals from two manifolds $\A$ and $\B$. 
For instance, suppose we are interested in the space of images, where
$\A$ and $\B$
comprise of translations of fixed template images $f(\t)$ and $g(\t)$,
where $\t$ denotes the 2D domain over which the image is defined. Then,
the signal of interest is an image of the form
\[
\c^* = \a^* + \b^* = f(\t+\theta_1) + g(\t + \theta_2),
\]
where $\theta_1$ and $\theta_2$ denote the unknown translation
parameters. The problem is to recover $(\a^*, \b^*)$, or equivalently
$(\theta_1, \theta_2)$, given compressive measurements $\z =
\bPhi(\a^*+\b^*)$.  

We demonstrate that SPIN offers an easy, efficient technique to
recover the component images. This example also demonstrates that SPIN
is robust to practical considerations such as noise. 
Figure~\ref{fig:circsq} displays the
results of SPIN recovery of a $64\times64$ image from very limited
measurements. The unknown image consists of the linear sum of
arbitrary translations of template images $f(\t)$ and $g(\t)$, that
are smoothed binary images on a black background of a white disk and a
white square, respectively. Further, the image has been contaminated
with significant Gaussian noise (SNR = 14dB) prior to measurement
(Fig.~\ref{fig:circsq}(a)). From Figs.~\ref{fig:circsq}(b) and \ref{fig:circsq}(c), we observe that SPIN is able to perfectly
recover the original  component signals from merely $M = 50$
random linear measurements. 

For guaranteed SPIN convergence, we require that
the manifolds $\A, \B$ are incoherent. Informally, the condition of incoherence on
the secants of $\A$ and $B$ is always valid when
the template images $f(\t), g(\t)$ are ``sufficiently''
distinct. This intuition is made precise using the bound in (\ref{eq:inch}).
More generally, we can state the following theoretical guarantee for
SPIN performance in the case of general higher-dimensional manifolds.
\begin{COROLLARY}[SPIN for pairs of manifolds]
\label{corr:man}
Let the entries of $\bPhi \in \real^{M \times N}$ be chosen from a
standard Gaussian probability distribution. Let $\A$ and  $\B$ be
$\epsilon$-incoherent compact submanifolds of $\real^N$ of dimensions
$K$ and $K'$ respectively.  
Let $\z = \bPhi(\a^* + \b^*)$, where $\a^* \in \A$ and $\b^* \in
\B$. Then, with high probability, SPIN
exactly recovers $(\a^*, \b^*)$ from $\z$, provided 
\begin{equation}
M = \bigo{ \left( K \log(C_\A N) + K' \log(C_\B N) \right)} .
\label{eq:pairs2}
\end{equation}
Here, $C_\A, C_\B$ are constants that depend only on certain intrinsic geometric parameters
(such as the volume) of $\A, \B$ respectively.
\end{COROLLARY}
{\em Proof.} It is easy to see that if the matrix $\bPhi$ satisfies the
RIP on the secants of the direct sum $\C = \A \oplus \B$, then SPIN recovery
follows from Theorem~\ref{thm:recovery}. We show that a randomized
construction of $\bPhi$ with number of rows specified by
(\ref{eq:pairs2}) satisfies the RIP on $\C$ with high probability.
Essentially, our proof combines the techniques used in Section 3.2
of~\cite{BaraniukWakin06} with Lemma 1 of~\cite{SPwithCS2010}. 

The manifold-embedding result in~\cite{BaraniukWakin06} is proved using two fundamental steps:
(i) careful construction of a {\em finite} subset $\R$ of points in
$\real^N$ that serves as a dense covering of the $K$-dimensional manifold of interest $\A$;
and (ii) application of the Johnson-Lindenstrauss lemma~\cite{JL} to
this finite set $\R$ to produce, with vanishingly low probability of
failure, a Gaussian measurement matrix
$\bPhi$ that satisfies the RIP on $\A$. Section 3.2.5
of~\cite{BaraniukWakin06} indicates that the cardinality of the the
finite set $\R$ can be upper bounded as
$$
\# \R \leq (C_\A N)^K,
$$
where $C_\A$ is a constant that depends only on the intrinsic geometry of
$\A$. 
However, in our setting we are interested in the direct sum of
manifolds; correspondingly, we can construct finite sets $\R_\A$ and
$\R_\B$ and apply Lemma 1 of~\cite{SPwithCS2010}, that specifies a
lower bound on the number of measurements required to preserve the
norms of linear sums of finite point sets:
\[
M \geq \bigo{\log( \R_\A \R_\B )} = \bigo{ \left( K \log(C_\A N) + K' \log(C_\B N) \right)} ,
\]
where $K, K'$ are the dimensions of $\A, \B$
respectively. Corollary~\ref{corr:man} follows. \qed

An important consideration in SPIN is the tractable computation of the projection $\P_\M(\x)$ given any $\x \in
\real^N$. For example, in the numerical example in Fig.\ \ref{fig:circsq},
 the operator $\P_\A(\x)$ onto the manifold $\A$
consists of running a matched filter between the template
$f(\t)$ and the input signal $\x$ and returning $f(\t + \widehat{\theta})$,
where the parameter value $\widehat{\theta}$ corresponds to the 2D
location of the maximum of the matched filter
response. This is very efficiently carried out in $\bigo{N \log N}$
operations using the Fast Fourier Transform (FFT). However, for more complex articulation manifolds, the exact
projection operation might not be tractable, whereas only approximate
numerical projections can be efficiently computed. In this case also, 
SPIN can recover the signal components $(\a^*, \b^*)$, but
with weaker convergence guarantees (Theorem~\ref{thm:approx}).

\subsection{Signals in impulsive noise}

\begin{figure*}[!t]
\centering
\begin{tabular}{ccc}
\includegraphics[width=0.3\textwidth]{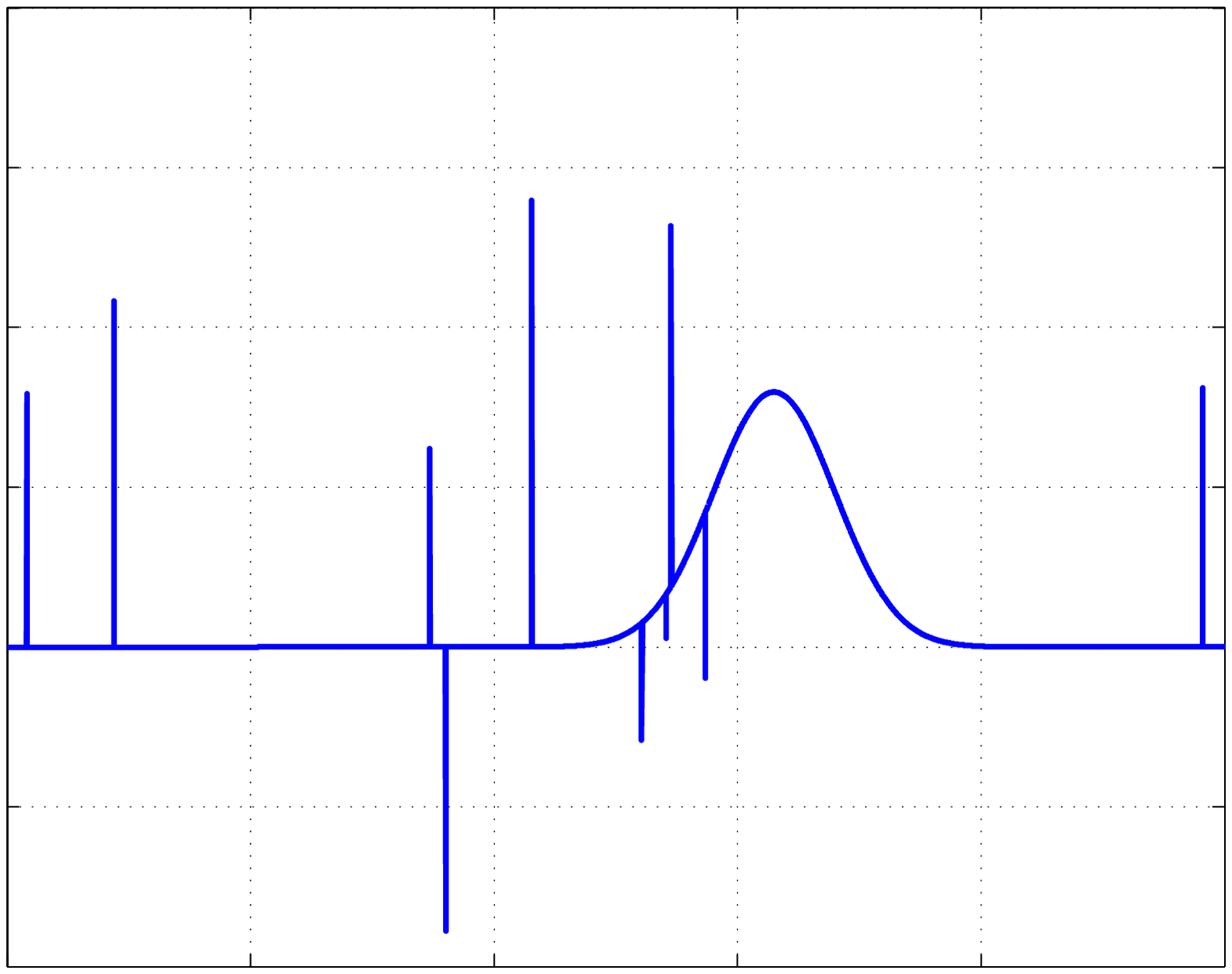} &
\includegraphics[width=0.3\textwidth]{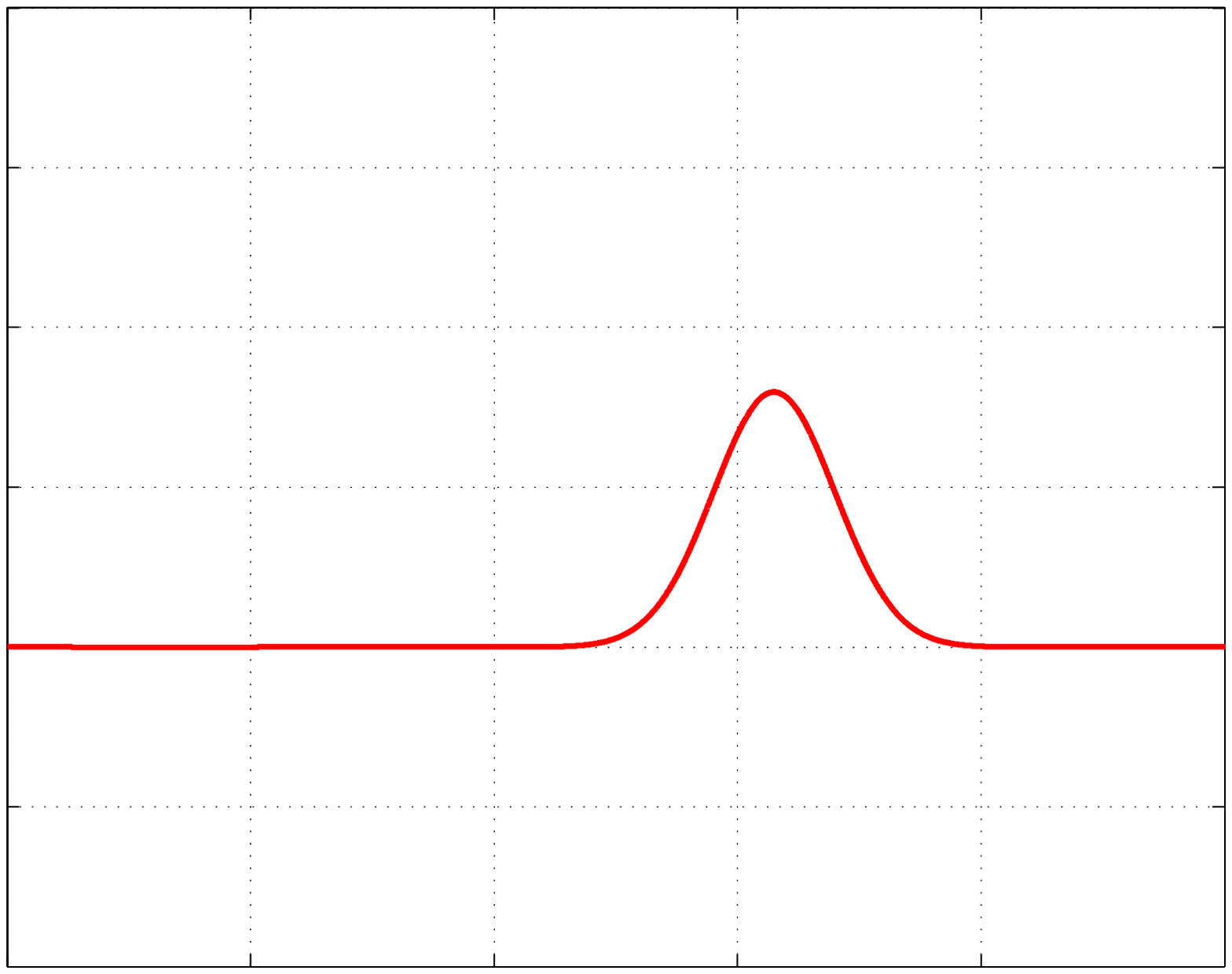} &
\includegraphics[width=0.3\textwidth]{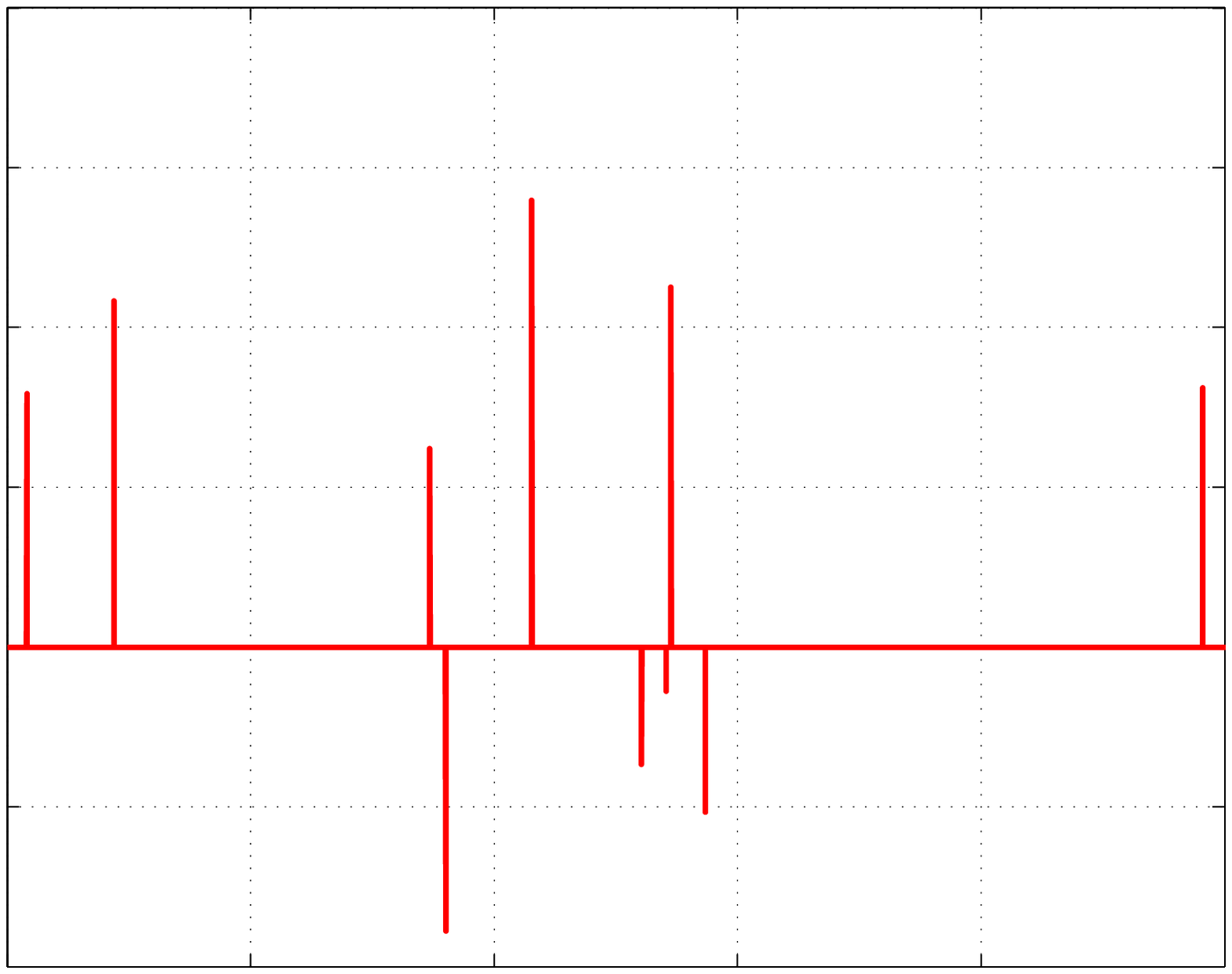} \\
{\small (a) Original signal} & {\small (b) Manifold component} &
{\small (c) Noise component}
\end{tabular}
\caption{{\small\sl SPIN recovery of a shifted Gaussian pulse from compressive
    measurements. The shift parameter of the pulse is unknown, and the signal is
    corrupted with $K'$-sparse, impulsive noise of unknown amplitudes and locations.
$N = 10000, K' = 10, M = 150.$ (a) Original signal. (b) Reconstructed
Gaussian pulse (Recovery SNR = 80.09 dB). (c) Estimated noise component. SPIN perfectly
reconstructs both components from just M/N = 1.5\% measurements.}
\label{fig:sparsegauss}
}
\end{figure*}

In some situations, the signal of interest $\x$ might be corrupted
with {\em impulsive noise} (or shot noise) prior to signal acquisition via
linear measurements. For example, consider Fig.\ \ref{fig:sparsegauss}(a), where the Gaussian
pulse is the signal of interest, and the spikes indicate the
undesirable noise. In this case, the linear observations are more
accurately modeled as:
$$
\z = \bPhi (\x + \n),~~\textrm{such that}~\x \in \M,  
$$
and $\n$ is a $K'$-sparse signal in the canonical basis. 
Therefore, SPIN can be used
to recover $\x$ from $\z$, provided that the
manifold $\M$ is incoherent with the set of sparse signals
$\Sigma_{K'}$ and $\bPhi$ satisfies the RIP on the direct sum $\M + \Sigma_{K'}$. 

Figure~\ref{fig:sparsegauss} displays the results of a numerical
experiment that illustrates the utility of SPIN in this setting. We
consider a manifold of signals of length $N=10000$ that consist of
shifts of a Gaussian pulse of fixed width $\g_0 \in \real^N$. The unknown signal $\x$ is
an element of this manifold $\M$, and is corrupted by $K'=10$ spikes
of unknown magnitudes and locations. This degraded signal is sampled
using $M=150$ random linear measurements to obtain an observation
vector $\z$. 

We apply SPIN to recover $\x$ from $\z$. The projection operator
$\P_\M(\cdot)$ consists of a matched filter with the template
pulse $\g_0$, while the projection operator $\P_{\Sigma_{K'}}(\cdot)$
simply returns the best $K'$-term approximation in the canonical
basis. Assuming that we have knowledge of the number of nonzeros in
the noise vector $\n$, we can use SPIN to reconstruct both $\x$ and $\n$. 
 We observe from Fig.\ \ref{fig:sparsegauss}(b) that SPIN recovers
the true signal $\x$ with near-perfect
accuracy. Further, this recovery is possible with only a small number $M = 150$
linear measurements of $\x$, which constitutes but a fraction of the
ambient dimension of the signal space. 

\begin{figure}[!t]
\centering
\includegraphics[width=0.45\linewidth]{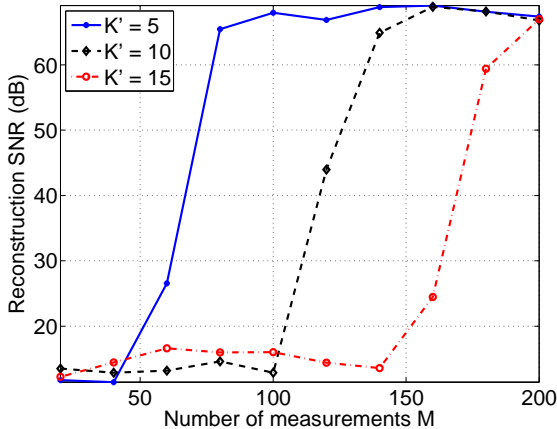} 
\caption{{\small\sl Monte Carlo simulation of SPIN signal recovery in
    impulsive noise 
   , averaged over 100 trials. In each trial, the measured
    signal is the sum of a randomly shifted Gaussian pulse
    and a random $K'$-sparse signal. SPIN can tolerate a higher number
    $K'$ nuisance impulses by increasing the number of measurements
    $M$. By Corollary~\ref{corr:man} this dependence of $M$ on $K'$
    can be shown to be {\em linear}.}
\label{fig:sparsegaussmc}
}
\qq
\end{figure}
Figure~\ref{fig:sparsegaussmc} plots the number of measurements $M$
vs.\ the signal reconstruction error (normalized relative to the
signal energy and plotted in dB). We observe that, by increasing $M$,
SPIN can tolerate an increased number $K'$ of nuisance
spikes. Further, by Corollary~\ref{corr:man}, we observe that this relationship between $M$ and $K'$ is in fact {\em
  linear}. This result can be extended to any situation where the signals of interest obey a
``hybrid'' model that is a mixture of a nonlinear manifold and the set
of sparse signals.


\section{Discussion}
\label{sec:conc}

We have proposed and rigorously analyzed an algorithm, which we dub
Successive Projections onto INcoherent Manifolds (SPIN), for the
recovery of a pair of signals given a small number of measurements of their
linear sum. For SPIN to guarantee signal recovery, we require two main
geometric criteria to hold: (i) the
component signals should arise from two disjoint manifolds that are in a specific
sense {\em incoherent}, and (ii) the linear measurement operator should
satisfy a {\em restricted isometry} criterion on the secants of the direct
sum of the two manifolds. The computational efficiency of SPIN is determined
by the tractability of the {\em projection operators} onto either
component manifold. We have presented indicative numerical experiments
demonstrating the utility of SPIN, but defer a thorough experimental
study of SPIN to future work.

{\bf\em Practical considerations.} SPIN is an iterative gradient
projection algorithm and requires as input parameters
the number of iterations $T$ and the gradient step size $\eta$. The
iteration count $T$ can be chosen using one of many commonly-used
stopping criteria. For example, convergence can be declared if the
norm of the error $\psi(\a_k, \b_k)$ at the $(k+1)$-th time step does not differ significantly
from the error at $k$-th time step. The choice of optimal step size
$\eta$ is more delicate. Theorem~\ref{thm:recovery} relates the step size
to the restricted isometry constant $\delta$ of $\bPhi$, but this constant is not
easy to calculate. In our preliminary findings, a step
size in the range $0.5 \leq \eta \leq 0.7$ consistently gave good
results. See~\cite{cevher} for a discussion on the choice of 
step size for hard thresholding methods.

In practical
scenarios, the signal of interest rarely belongs exactly to a low-dimensional
submanifold $\M$ of the ambient space, but is only well-approximated
by $\M$. Interestingly, in such situations the effect of this mismatch can be
studied using the concept of $\gamma$-approximate projections
(\ref{eq:projapprox}). Theorem~\ref{thm:approx} rigorously demonstrates that SPIN is robust to
such approximations. Further, our main result (Theorem~\ref{thm:recovery}) indicates that SPIN is
stable with respect to inaccurate measurements, owing to the fact that
the reconstruction error is bounded by a constant times the norm of
the measurement noise vector $\e$.

{\bf\em More than two manifolds.} For clarity and brevity, we have
focused our attention on signals belonging to the direct sum of two signal manifolds. However, SPIN (and its
accompanying proof mechanism) can be conceptually extended to sums of
any $Q$ manifolds. In such a scenario, the conditions of convergence
of SPIN would require that the component manifolds are $Q$-wise incoherent, and the measurement
operator $\bPhi$ satisfies a restricted isometry on the $Q$-wise direct sum of
the component manifolds.

{\bf\em Connections to matrix recovery.} An intriguing open question is whether SPIN (or a similar first-order projected
gradient algorithm) is applicable to situations where either of the
component manifolds is the set of low-rank matrices. The problem of
reconstructing, from affine measurements, matrices that are a sum of
low-rank and sparse matrices has attracted significant attention in the
recent
literature~\cite{candes2009robust,drewnips,chandrasekaran2009sparse}. The key
stumbling block is that the manifold of low-rank matrices is {\em not incoherent}
with the manifold of sparse matrices; indeed, the two manifolds share
a nontrivial intersection (i.e., there exist low rank matrices that
are also sparse, and vice versa). Phenomena such as these make the
analysis of SPIN (or similar algorithms) quite challenging, and it
may be that higher-order techniques will be needed for signal recovery. 

{{
\bibliography{csbib}
\bibliographystyle{IEEEbib}
}}

\end{document}